\documentclass[12pt,titlepage,letterpaper]{utarticle}

\usepackage{amsmath}
\usepackage{amsfonts}
\usepackage{amssymb}
\usepackage{amsthm}
\usepackage{xparse}
\usepackage{mathtools}
\usepackage{graphicx}
\usepackage{color}
\usepackage[utf8x]{inputenc}
\usepackage[titletoc,title]{appendix}
\usepackage{hyperref}
\usepackage{longtable}

\definecolor{aqua}{rgb}{0, 1.0, 1.0}
\definecolor{fuschia}{rgb}{1.0, 0, 1.0}
\definecolor{gray}{rgb}{0.502, 0.502, 0.502}
\definecolor{lime}{rgb}{0, 1.0, 0}
\definecolor{maroon}{rgb}{0.502, 0, 0}
\definecolor{navy}{rgb}{0, 0, 0.502}
\definecolor{olive}{rgb}{0.502, 0.502, 0}
\definecolor{purple}{rgb}{0.502, 0, 0.502}
\definecolor{silver}{rgb}{0.753, 0.753, 0.753}
\definecolor{teal}{rgb}{0, 0.502, 0.502}

\makeatletter
\newdimen\itex@wd%
\newdimen\itex@dp%
\newdimen\itex@thd%
\def\itexspace#1#2#3{\itex@wd=#3em%
\itex@wd=0.1\itex@wd%
\itex@dp=#2ex%
\itex@dp=0.1\itex@dp%
\itex@thd=#1ex%
\itex@thd=0.1\itex@thd%
\advance\itex@thd\the\itex@dp%
\makebox[\the\itex@wd]{\rule[-\the\itex@dp]{0cm}{\the\itex@thd}}}
\makeatother

\makeatletter
\newif\if@sup
\newtoks\@sups
\def\append@sup#1{\edef\act{\noexpand\@sups={\the\@sups #1}}\act}%
\def\reset@sup{\@supfalse\@sups={}}%
\def\mk@scripts#1#2{\if #2/ \if@sup ^{\the\@sups}\fi \else%
  \ifx #1_ \if@sup ^{\the\@sups}\reset@sup \fi {}_{#2}%
  \else \append@sup#2 \@suptrue \fi%
  \expandafter\mk@scripts\fi}
\def\tensor#1#2{\reset@sup#1\mk@scripts#2_/}
\def\multiscripts#1#2#3{\reset@sup{}\mk@scripts#1_/#2%
  \reset@sup\mk@scripts#3_/}
\makeatother

\makeatletter
\newbox\slashbox \setbox\slashbox=\hbox{$/$}
\def\itex@pslash#1{\setbox\@tempboxa=\hbox{$#1$}
  \@tempdima=0.5\wd\slashbox \advance\@tempdima 0.5\wd\@tempboxa
  \copy\slashbox \kern-\@tempdima \box\@tempboxa}
\def\slash{\protect\itex@pslash}
\makeatother

\def\clap#1{\hbox to 0pt{\hss#1\hss}}

\let\oldroot\root
\def\root#1#2{\oldroot #1 \of{#2}}
\renewcommand{\sqrt}[2][]{\oldroot #1 \of{#2}}

\DeclareSymbolFont{symbolsC}{U}{txsyc}{m}{n}
\SetSymbolFont{symbolsC}{bold}{U}{txsyc}{bx}{n}
\DeclareFontSubstitution{U}{txsyc}{m}{n}

\DeclareSymbolFont{stmry}{U}{stmry}{m}{n}
\SetSymbolFont{stmry}{bold}{U}{stmry}{b}{n}

\DeclareFontFamily{OMX}{MnSymbolE}{}
\DeclareSymbolFont{mnomx}{OMX}{MnSymbolE}{m}{n}
\SetSymbolFont{mnomx}{bold}{OMX}{MnSymbolE}{b}{n}
\DeclareFontShape{OMX}{MnSymbolE}{m}{n}{
    <-6>  MnSymbolE5
   <6-7>  MnSymbolE6
   <7-8>  MnSymbolE7
   <8-9>  MnSymbolE8
   <9-10> MnSymbolE9
  <10-12> MnSymbolE10
  <12->   MnSymbolE12}{}

\makeatletter
\def\re@DeclareMathSymbol#1#2#3#4{%
    \let#1=\undefined
    \DeclareMathSymbol{#1}{#2}{#3}{#4}}
\re@DeclareMathSymbol{\neArrow}{\mathrel}{symbolsC}{116}
\re@DeclareMathSymbol{\neArr}{\mathrel}{symbolsC}{116}
\re@DeclareMathSymbol{\seArrow}{\mathrel}{symbolsC}{117}
\re@DeclareMathSymbol{\seArr}{\mathrel}{symbolsC}{117}
\re@DeclareMathSymbol{\nwArrow}{\mathrel}{symbolsC}{118}
\re@DeclareMathSymbol{\nwArr}{\mathrel}{symbolsC}{118}
\re@DeclareMathSymbol{\swArrow}{\mathrel}{symbolsC}{119}
\re@DeclareMathSymbol{\swArr}{\mathrel}{symbolsC}{119}
\re@DeclareMathSymbol{\nequiv}{\mathrel}{symbolsC}{46}
\re@DeclareMathSymbol{\Perp}{\mathrel}{symbolsC}{121}
\re@DeclareMathSymbol{\Vbar}{\mathrel}{symbolsC}{121}
\re@DeclareMathSymbol{\sslash}{\mathrel}{stmry}{12}
\re@DeclareMathSymbol{\bigsqcap}{\mathop}{stmry}{"64}
\re@DeclareMathSymbol{\biginterleave}{\mathop}{stmry}{"6}
\re@DeclareMathSymbol{\invamp}{\mathrel}{symbolsC}{77}
\re@DeclareMathSymbol{\parr}{\mathrel}{symbolsC}{77}
\makeatother

\makeatletter
\def\Decl@Mn@Delim#1#2#3#4{%
  \if\relax\noexpand#1%
    \let#1\undefined
  \fi
  \DeclareMathDelimiter{#1}{#2}{#3}{#4}{#3}{#4}}
\def\Decl@Mn@Open#1#2#3{\Decl@Mn@Delim{#1}{\mathopen}{#2}{#3}}
\def\Decl@Mn@Close#1#2#3{\Decl@Mn@Delim{#1}{\mathclose}{#2}{#3}}
\Decl@Mn@Open{\llangle}{mnomx}{'164}
\Decl@Mn@Close{\rrangle}{mnomx}{'171}
\Decl@Mn@Open{\lmoustache}{mnomx}{'245}
\Decl@Mn@Close{\rmoustache}{mnomx}{'244}
\makeatother

\makeatletter
\DeclareRobustCommand\widecheck[1]{{\mathpalette\@widecheck{#1}}}
\def\@widecheck#1#2{%
    \setbox\z@\hbox{\m@th$#1#2$}%
    \setbox\tw@\hbox{\m@th$#1%
       \widehat{%
          \vrule\@width\z@\@height\ht\z@
          \vrule\@height\z@\@width\wd\z@}$}%
    \dp\tw@-\ht\z@
    \@tempdima\ht\z@ \advance\@tempdima2\ht\tw@ \divide\@tempdima\thr@@
    \setbox\tw@\hbox{%
       \raise\@tempdima\hbox{\scalebox{1}[-1]{\lower\@tempdima\box
\tw@}}}%
    {\ooalign{\box\tw@ \cr \box\z@}}}
\makeatother

\makeatletter
\NewDocumentCommand\mathraisebox{moom}{%
\IfNoValueTF{#2}{\def\@temp##1##2{\raisebox{#1}{$\m@th##1##2$}}}{%
\IfNoValueTF{#3}{\def\@temp##1##2{\raisebox{#1}[#2]{$\m@th##1##2$}}%
}{\def\@temp##1##2{\raisebox{#1}[#2][#3]{$\m@th##1##2$}}}}%
\mathpalette\@temp{#4}}
\makeatletter

\makeatletter
\def\udots{\mathinner{\mkern2mu\raise\p@\hbox{.}
\mkern2mu\raise4\p@\hbox{.}\mkern1mu
\raise7\p@\vbox{\kern7\p@\hbox{.}}\mkern1mu}}
\makeatother

\theoremstyle{plain}

\theoremstyle{definition}

\theoremstyle{remark}

\begin{document}


\preprint{
UTTG--XX--15\\
TCC--XXX--15\\
ICTP--SAIFR/2015--XXX\\
}

\title{A Family of $4D$ $\mathcal{N}=2$ Interacting SCFTs from the Twisted $A_{2N}$ Series}

\author{Oscar Chacaltana
    \address{
    ICTP South American Institute for\\ Fundamental Research,\\
    Instituto de F\'isica Te\'orica,\\Universidade Estadual Paulista,\\
    01140-070 S\~{a}o Paulo, SP, Brazil\\
    {~}\\
    \email{chacaltana@ift.unesp.br}\\
    },
    Jacques Distler ${}^\mathrm{b}$ and Anderson Trimm
     \address{
      Theory Group and\\
      Texas Cosmology Center\\
      Department of Physics,\\
      University of Texas at Austin,\\
      Austin, TX 78712, USA \\
      {~}\\
      \email{distler@golem.ph.utexas.edu}\\
      \email{atrimm@physics.utexas.edu}
      }
}
\date{December 28, 2014}

\Abstract{
We find an infinite family of $4D$ $\mathcal{N}=2$ interacting superconformal field theories which enter the description of the strong-coupling limit of  $SU(2N+1)$ gauge theories with hypermultiplets in the $\wedge^2(\square)+\text{Sym}^2(\square)$ . These theories arise from the compactification of the $6D$ $(2,0)$ theory of type $A_{2N}$ on a sphere with two full twisted punctures and one minimal untwisted puncture. For $N=1$, this theory is the ``new" rank-1 SCFT with $\Delta(u)=3$ of Argyres and Wittig. Using the superconformal index, we finally pin down the properties of this theory.
}

\maketitle

\thispagestyle{empty}
\tableofcontents
\vfill
\newpage
\setcounter{page}{1}

\section{Introduction}\label{introduction}

In \cite{Gaiotto:2009hg,Gaiotto:2009we}, it was shown that one can construct and systematically understand S-duality for a large family of $4D$ $\mathcal{N}=2$ superconformal field theories by realizing them as the compactification of a $6D$ $(2,0)$ theory on a punctured Riemann surface, $C$. This construction has lead to the discovery of a large number of new $\mathcal{N}=2$ interacting SCFTs in four dimensions, which are isolated fixed points of the renormalization group with no known Lagrangian description \cite{Gaiotto:2009we,Chacaltana:2010ks,Chacaltana:2011ze,Chacaltana:2012ch,Chacaltana:2013oka,Chacaltana:2014jba,Chacaltana:2014ica}. A subset of these are especially interesting as (when some subgroup of their global symmetry group is weakly gauged) they describe the strong-coupling limits of ordinary $\mathcal{N}=2$ theories of vector and hypermultiplets.

In this paper, we consider the strong-coupling limit of $SU(2N+1)$ gauge theory with hypermultiplets in the $1\Bigl(\begin{matrix} \includegraphics[width=9pt]{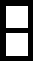}\end{matrix}\Bigr)+1\bigl( \includegraphics[width=17pt]{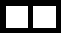}\bigr)$. We find the following S-duality for this theory

\begin{equation}
SU(2N+1)+1\Bigl(\begin{matrix} \includegraphics[width=9pt]{antisym}\end{matrix}\Bigr)+1\bigl( \includegraphics[width=17pt]{sym}\bigr) \simeq Sp(N) + R_{2,2N}
\label{Sduality}\end{equation}
where $R_{2,2N}$, $N \geq 1$, belongs to a family of interacting SCFTs with the following graded Coulomb branch dimensions, trace anomaly coefficients, and global symmetry:

\bigskip
\begin{tabular}{|c|c|c|c|}
\hline
&$\{d_2,d_3,d_4,d_5,\dots,d_{2N},d_{2N+1}\}$&$(a,c)$&$G_\text{global}$\\
\hline 
$\vphantom{\biggl(}R_{2,2N}$&$\{0,1,0,1,\dots, 0, 1\}$&$(\frac{1+19N+14N^2}{24},\frac{1+10N+8N^2}{12})$&$Sp(2N)_{2N+2} \times U(1)$\\
\hline
\end{tabular}
\bigskip

The proposed duality \eqref{Sduality} is analogous to the duality
$$
  SU(2N) + 1\Bigl(\begin{matrix} \includegraphics[width=9pt]{antisym}\end{matrix}\Bigr)+1\bigl( \includegraphics[width=17pt]{sym}\bigr) \simeq Spin(2N+1) + R_{2,2N-1}
$$
discussed\footnote{The $R_{2,2N-1}$ series of SCFTs, which have global symmetry $Spin(4N+2)_{4N−2}\times U(1)$ also play a role in the dualities \cite{Chacaltana:2010ks}
\begin{equation*}
\begin{split}
SU(2N-1) +4\bigl(\square\bigr)+2\Bigl(\begin{matrix}\includegraphics[width=9pt]{antisym}\end{matrix}\Bigr)
&\simeq Sp(N-1) + 1\bigl(\square\bigr)
+ R_{2,2N-1}\\
SU(2N) +4\bigl(\square\bigr)+2\Bigl(\begin{matrix}\includegraphics[width=9pt]{antisym}\end{matrix}\Bigr)
&\simeq Sp(N) + 3\bigl(\square\bigr) + R_{2,2N-1}
\end{split}
\end{equation*}} in \S 3.5.4 of \cite{Chacaltana:2012zy}. However the $R_{2,2N}$ series of SCFTs is \emph{new}.

The strong-coupling limit of $SU(3)+ 1(3) +1( 6)$ was considered by Argyres and Wittig in \cite{Argyres:2007tq}. They conjectured that this theory is dual to an $SU(2)$ gauge theory with $n \leq 3$ fundamental hypermultiplets, coupled to a new rank 1 interacting SCFT. This S-duality alone does not fix $n$, and the properties of this theory have remained only partially-known. In \cite{Chacaltana:2012ch}, we gave a $6D$ realization of this S-duality, which is given by taking $N=1$ in the figure below. From this construction, the properties of the holomorphic moment map operators for the flavor symmetry of the two twisted punctures imply that $n=0$ \cite{Chacaltana:2012ch,Gaiotto:2011xs}. In the following, we will provide independent evidence that $n=0$ using the superconformal index. In addition, we will use the index to determine the enhanced global symmetry of the SCFT, and generalize this duality to arbitrary $N$.

The main tool in our analysis is the Hall-Littlewood limit of the superconformal index \cite{Kinney:2005ej,Gadde:2011uv, Rastelli:2014jja}. In this limit, the superconformal index is equivalent to the Coulomb branch Hilbert series of the $3D$ mirror of the $(2,0)$ theory on $C \times S^1$ \cite{Benini:2010uu, Gaiotto:2012uq}. The $3D$ interpretation allows us to easily obtain the formula for the index of a fixture with twisted $A_{2N}$ punctures, following \cite{Cremonesi:2014kwa,Cremonesi:2014vla}. The result is \eqref{SCIhigher}.

To construct the $R_{2,2N}$ theories, we consider compactifications of the $A_{2N}$ $(2,0)$ theory in the presence of $\mathbb{Z}_2$ outer-automorphism twists\footnote{Outer-automorphism twists of type $A_{2N-1}, D_N$, and $E_6$ have been studied in \cite{Chacaltana:2012ch,Chacaltana:2013oka, Tachikawa:2009rb,Tachikawa:2010vg, Lemos:2012ph,Chacaltana:2014abc,Chacaltana:2014def}.}. As discussed in \cite{Tachikawa:2011ch}, these twists are particularly subtle, and we do not yet understand them well enough to attempt a systematic classification of the $4D$ theories which arise in this way. Nevertheless, our current level of understanding is sufficient to use them for our purposes here.

\section{S-duality of $SU(2N+1)+\wedge^2(\square)+\text{Sym}^2(\square)$
}\label{sduality_of_}

The $4D$ $\mathcal{N}=2$ $SU(2N+1)$ gauge theory with hypermultiplets in the $1\Bigl(\begin{matrix} \includegraphics[width=9pt]{antisym}\end{matrix}\Bigr)+1\bigl( \includegraphics[width=17pt]{sym}\bigr)$ can be constructed as follows \cite{Chacaltana:2012ch}. The $A_{2N}$ $(2,0)$ theory compactified on a fixture with two full punctures and one minimal puncture gives rise to a free bifundamental hypermultiplet of $SU(2N+1)$. One can gauge the diagonal $SU(2N+1)$ flavor symmetry of the two full punctures by connecting them by a cylinder with a $\mathbb{Z}_2$ twist line around it, to obtain a one-punctured torus, with a twist line around the $a$-cycle. This is shown on the left in the figure below. The $\mathbb{Z}_2$-twist line acts as complex conjugation on one of the $SU(2N+1)$ factors, giving rise to hypermultiplets in the tensor product representation $\square\otimes\square= \begin{matrix} \includegraphics[width=9pt]{antisym}\end{matrix}+ \includegraphics[width=17pt]{sym}$ .

The S-dual theory is obtained by exchanging the $a$- and $b$-cycles of the torus. The resulting theory can be seen to arise by compactifying the $A_{2N}$ theory on a fixture with a minimal untwisted puncture and two full twisted punctures, and connecting the two twisted punctures by a cylinder with a twist line running along it. This is shown on the right in the figure below. We expect this to give rise to an $Sp(N)$ gauge theory coupled to an interacting SCFT, possibly with $n$ additional hypermultiplets.

\begin{displaymath}
 \includegraphics[width=412.5pt]{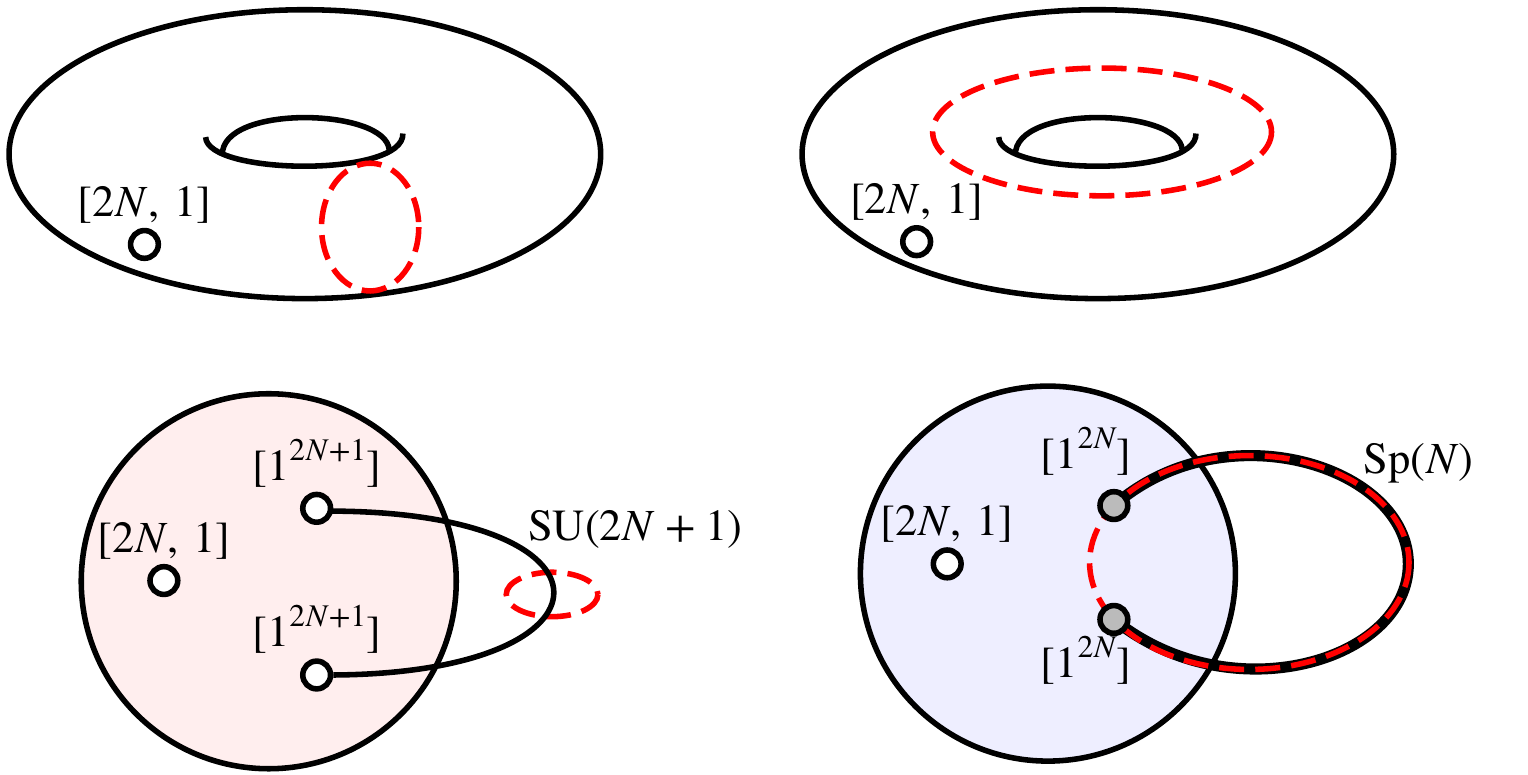}
\end{displaymath}
In the following, we will use the superconformal index to show that $n=0$, and that the manifest $Sp(N)_{2N+2} \times Sp(N)_{2N+2} \times U(1)$ global symmetry of the interacting SCFT is enhanced to $Sp(2N)_{2N+2} \times U(1)$. The graded Coulomb branch dimensions and trace anomaly coefficients of the SCFT then follow from S-duality.

\section{The Argyres-Wittig SCFT}\label{the_argyreswittig_scft}
\subsection{Global symmetry enhancement}\label{global_symmetry_enhancement}

We begin by considering the case of $N=1$. The fixture on the right in the figure above then has a manifest $SU(2)_{4} \times SU(2)_{4} \times U(1)$ subgroup of its global symmetry group, which could be enhanced. We will now use the superconformal index to show that this fixture contains an interacting SCFT with no additional free hypermultiplets, and that the global symmetry of the SCFT is enhanced to $Sp(2)_4 \times U(1)$.

We assume that the superconformal index for this fixture has the same general structure one always finds for a 3-punctured sphere, namely\footnote{The various parts of this formula have been explained many times in the references above. See, e.g., \cite{Mekareeya:2012tn}.} 

\begin{equation}
\mathcal{I}(\mathbf{a}_i;\tau)=\sum_{\lambda=0}^\infty\frac{\prod_{i=1}^2\mathcal{K}(\mathbf{a}_i)P^{\lambda}_{SU(2)}(\mathbf{a}_i;\tau)\mathcal{K}(\mathbf{a}_3)P^{(2\lambda,\lambda)}_{SU(3)}(\mathbf{a}_3;\tau)}{\mathcal{K}_\rho P^{(2\lambda,\lambda)}_{SU(3)}(\mathbf{a}_\rho;\tau)}
\label{SCIfixt}\end{equation}
Assigning fugacities to each puncture, this becomes

\begin{displaymath}
\begin{split}
\mathcal{I}(\mathbf{a}_i;\tau)&=\frac{(1-\tau^2)(1-\tau^4)(1-\tau^6)}{(1-\tau^2)^3(1-\tau^2a_1^{\pm 2})(1-\tau^2a_2^{\pm 2})(1-\tau^3 a_3^{\pm 3})(1-\tau^4)} \\
&\times \left(1+\tau^2+\sum_{\lambda=1}^\infty \frac{P^\lambda_{SU(2)}(a_1,a_1^{-1};\tau)P^\lambda_{SU(2)}(a_2,a_2^{-1};\tau)P^{(2\lambda,\lambda)}_{SU(3)}(a_3 \tau ,a_3 \tau^{-1},a_3^{-2};\tau)}{P^{(2\lambda,\lambda)}_{SU(3)}(\tau^2 ,1,\tau^{-2};\tau)}\right) \\
&=1+(1+\chi^{\mathbf{10}}_{Sp(2)}(a_1,a_2))\tau^2+\dots
\end{split}
\end{displaymath}
where

\begin{equation}
\chi^{\mathbf{10}}_{Sp(2)}(a_1,a_2)=\chi^{\mathbf{3}}_{SU(2)}(a_1)+\chi^{\mathbf{3}}_{SU(2)}(a_2)+\chi^{\mathbf{2}}_{SU(2)}(a_1)\chi^{\mathbf{2}}_{SU(2)}(a_2).
\label{Sp2}\end{equation}
The coefficient of $\tau$ is zero, indicating that the fixture contains no additional free hypermultiplets. The coefficient of $\tau^2$ is therefore the character of the adjoint representation of the global symmetry group of the SCFT, which is enhanced to $Sp(2) \times U(1)$. Since the embedding \eqref{Sp2} has index 1, the level of $Sp(2)$ is $k=4$.

Setting $a_1=a_2=a_3=1$, we can sum \eqref{SCIfixt} to obtain

\begin{displaymath}
\mathcal{I}=\frac{1+2\tau+8\tau^2+20\tau^3+41\tau^4+62\tau^5+87\tau^6+96\tau^7+\cdots (\text{palindrome}) \cdots +\tau^{14}}{(1-\tau)^8(1+\tau)^6(1+\tau+\tau^2)^4}.
\end{displaymath}
This expression takes the expected form (see, e.g., \cite{Hanany:2012dm}), and the order of the pole at $\tau=1$ gives the complex dimension of the $4D$ Higgs branch, which agrees with the answer obtained from S-duality \cite{Chacaltana:2010ks,Chacaltana:2012zy}

\begin{displaymath}
\text{dim}_{\mathbb{C}}\mathcal{H}=48(c-a)=48\left(\frac{19}{12}-\frac{17}{12}\right)=8.
\end{displaymath}

\subsection{Argyres-Wittig duality}\label{argyreswittig_duality}

As a further check on the validity of our computation, we compare the index on both sides of the S-duality

\begin{displaymath}
SU(3)+1(3)+1(6) \simeq SU(2) + R_{2,2}
\end{displaymath}
The $SU(3)$ theory is a Lagrangian theory, so its index is computed by the matrix integral \footnote{For simplicity, we set the fugacities of the $U(1)^2$ flavor symmetry to 1.} 

\begin{displaymath}
\begin{split}
\mathcal{I}&=\oint [d\mathbf{z}]_{SU(3)}PE[-\tau^2 \chi^{\mathbf{8}}_{SU(3)}(z_1,z_2)]PE[\tau \chi^{\mathbf{3}}_{SU(3)}(z_1,z_2)]PE[\tau \chi^{\mathbf{\overline{3}}}_{SU(3)}(z_1,z_2)]\\ 
&\times PE[\tau \chi^{\mathbf{6}}_{SU(3)}(z_1,z_2)]PE[\tau \chi^{\mathbf{\overline{6}}}_{SU(3)}(z_1,z_2)] \\
&=\frac{1}{3!}\oint \frac{dz_1dz_2}{(2\pi i)^2 z_1 z_2}(1-(z_1/z_2)^\pm)(1-z_1^{\pm 2}z_2^\pm)(1-z_1^\pm z_2^{\pm 2}) \\
&\times \frac{(1-\tau^2)^2(1-\tau^2(z_1/z_2)^\pm)(1-\tau^2z_1^{\pm 2}z_2^\pm)(1-\tau^2z_1^\pm z_2^{\pm 2})}{(1-\tau z_1^\pm)^2(1-\tau z_2^\pm)^2(1-\tau(z_1z_2)^\pm)^2(1-\tau z_1^{\pm 2})(1-\tau z_2^{\pm 2})(1-\tau (z_1 z_2)^{\pm 2})}.
\end{split}
\end{displaymath}
Summing the residues at the poles inside the unit circle (taking $|z_1|=|z_2|=1,\tau<1$), we arrive at

\begin{equation}
\mathcal{I}=\frac{1+\tau^2+2\tau^3+\tau^4}{(1-\tau)^3(1+\tau)(1+\tau+\tau^2)^2}.
\label{answer}\end{equation}
On the $SU(2)$ side of the duality, the index is given by

\begin{displaymath}
\mathcal{I}=\frac{1}{2!}\oint \frac{da}{2\pi i a}(1-a^{\pm 2})\mathcal{I}^V_{SU(2)}(a)\mathcal{I}_{R_{2,2}}(a,a^{-1},1),
\end{displaymath}
where $\mathcal{I}^V_{SU(2)}(a)$ is the index of a free $SU(2)$ vector multiplet. The integrand has poles inside the unit circle at $a=0,\pm \tau, \pm \tau^{3/2}$. Summing the residues, we reproduce \eqref{answer}. This gives strong evidence that \eqref{SCIfixt} indeed gives the index of the Argyres-Wittig SCFT.

\subsection{Chiral ring}\label{chiral_ring}
We can study the Hall-Littlewood chiral ring \cite{Beem:2014rza} of the Argyres-Wittig SCFT by taking the plethystic log of \eqref{SCIfixt}:

\begin{displaymath}
\begin{split}
PL[\mathcal{I}]&=\tau^2(\chi^{\mathbf{10}}_{Sp(2)}(a_1,a_2)+1)+\tau^3(\chi^{\mathbf{5}}_{Sp(2)}(a_1,a_2)a_3^3+\chi^{\mathbf{5}}_{Sp(2)}(a_1,a_2)a_3^{-3})-\tau^4(\chi^{\mathbf{5}}_{Sp(2)}(a_1,a_2)+1)\\
&-\tau^5((\chi^{\mathbf{10}}_{Sp(2)}(a_1,a_2)+\chi^{\mathbf{5}}_{Sp(2)}(a_1,a_2))a_3^3+(\chi^{\mathbf{10}}_{Sp(2)}(a_1,a_2)+\chi^{\mathbf{5}}_{Sp(2)}(a_1,a_2))a_3^{-3})-\dots
\end{split}
\end{displaymath}
where $\chi^{\mathbf{5}}(a_1,a_2)=1+\chi^{\mathbf{2}}_{SU(2)}(a_1)\chi^{\mathbf{2}}_{SU(2)}(a_2)$.

From this expression, we see that the chiral ring has generators at order 2 in the $\mathbf{10}_0+\mathbf{1}_0$ of $Sp(2) \times U(1)$ (which are the holomorphic moment map operators for the global symmetry), as well as order 3 generators in the $\mathbf{5}_3+\mathbf{5}_{-3}$. These generators are subject to relations at order 4 in the $\mathbf{5}_0+\mathbf{1}_0$, as well as higher order relations.

We can understand the two relations at order 4 as follows. In \cite{Beem:2014rza, Beem:2013sza}, it was shown that any $4D$ $\mathcal{N}=2$ SCFT contains families of protected operators whose correlation functions possess the structure of a $2D$ chiral algebra. The existence of the chiral algebra structure leads to additional unitarity bounds on central charges in the $4D$ theory, and saturation of these bounds was shown to follow from Higgs branch chiral ring relations. The operators counted by the Hall-Littlewood superconformal index fall into this class of protected operators, and from table 3 in \cite{Beem:2013sza}, we see that the level of the $Sp(2)$ factor in the global symmetry group saturates the unitarity bound, which follows from an order 4 relation in the $\mathbf{5}$. It was also shown in \cite{Beem:2013sza} that an order 4 relation in the $\mathbf{1}$ implies that the $2D$ chiral algebra has a stress tensor given by the Sugawara construction, with central charge:

\begin{equation}\label{sugawara}
c_{2D}=\frac{k_{2D}\text{dim }G_F}{k_{2D}+h^\vee}
\end{equation}
where the $2D$ central charges are related to those in $4D$ by $c_{2D}=-12c_{4D}$, $k_{2D}=-\frac{k_{4D}}{2}$.

Indeed, we find that

\begin{displaymath}
c_{4D}=-\frac{1}{12}\left(\frac{(-2)(10)}{-2+1}+1\right)=\frac{19}{12}.
\end{displaymath}

\section{Higher $N$}\label{higher_N}

We now consider the compactification of the $6D$ $(2,0)$ theory of type $A_{2N}$ on

\begin{displaymath}
 \includegraphics[width=103pt]{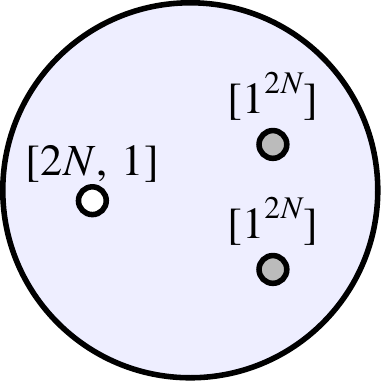}
\end{displaymath}
and use the superconformal index to argue that these fixtures describe the $R_{2,2N}$ family of interacting SCFTs, with no additional hypermultiplets.

The Hall-Littlewood index is given by

\begin{equation}
\mathcal{I}(\mathbf{x}_i;\tau)=\sum_{\lambda_1\geq \lambda_2 \geq \dots \geq \lambda_N \geq 0}\frac{\prod_{i=1}^2\mathcal{K}(\mathbf{x}_i)P^{(\lambda_1,\dots,\lambda_N)}_{Sp(N)}(\mathbf{x}_i;\tau)\mathcal{K}(\mathbf{x}_3)P^{(\lambda_1',\dots,\lambda_N')}_{SU(2N+1)}(\mathbf{x}_3;\tau)}{\mathcal{K}_\rho P^{(\lambda_1',\dots,\lambda_N')}_{SU(2N+1)}(\mathbf{x}_\rho;\tau)}
\label{SCIhigher}\end{equation}
where the sum runs over integers $\lambda_i$ labeling $Sp(N)$ irreps, and the corresponding $SU(2N+1)$ representations are given by $(\lambda_1',\dots,\lambda_N')=(2\lambda_1,\lambda_1+\lambda_2,\dots, \lambda_1+\lambda_N,\lambda_1,\lambda_1-\lambda_{N},\dots,\lambda_1-\lambda_2)$.

We can characterize these fixtures by the leading power of $\tau$ in the expansion of \eqref{SCIhigher}, as follows \cite{Gaiotto:2012uq}: The term coming from taking $(\lambda_1,\dots,\lambda_N)=(0,\dots,0)$ in the sum is 

\begin{displaymath}
1+\tau^2(\chi^{\mathbf{N(2N+1)}}_{Sp(N)}(\mathbf{x}_1)+\chi^{\mathbf{N(2N+1)}}_{Sp(N)}(\mathbf{x}_2)+1)+\mathcal{O}(\tau^4),
\end{displaymath}
encoding the manifest $Sp(N)_{2N+2} \times Sp(N)_{2N+2} \times U(1)$ global symmetry. This global symmetry is enhanced if there are additional terms at order $\tau^2$ coming from the sum over non-trivial representations. If there are also terms at order $\tau$, then the fixture contains free hypermultiplets along with the interacting SCFT.

Following \cite{Gaiotto:2012uq}, we give the leading behavior of \eqref{SCIhigher} in terms of a vector $v$, with components

\begin{equation}
v_i=(2N+2-2i)+\sum_{\ell=1}^3 d_i^{(\ell)}
\label{v}\end{equation}
where the first term is the leading power of $\tau$ from the denominator in \eqref{SCIhigher}, and the second term is the leading power of $\tau$ from the punctures. For the full twisted punctures, $d_i^{(\ell)}=0$ for all $i$, while for the minimal untwisted puncture we have

\begin{displaymath}
d_i^{(3)}=\begin{cases} &2i-2N-1,  \hspace{0.5cm} 1 \leq i \leq N, \\ &0,  \hspace{2.5cm}  i=N+1, \\ &2i-2N-3,  \hspace{0.5cm}  N+2 \leq i \leq 2N.\end{cases}
\end{displaymath}
Equation \eqref{v} then gives

\begin{displaymath}
v_i=\begin{cases} &1,  \hspace{0.9cm}  1 \leq i \leq N, \\ &0, \hspace{0.9cm} i=N+1, \\ &-1,  \hspace{0.5cm}  N+2 \leq i \leq 2N.\end{cases}
\end{displaymath}
The leading power in $\tau$ of \eqref{SCIhigher} is therefore given by

\begin{equation}
\begin{split}
v \cdot \lambda'&=(N+1)\lambda_1+\lambda_2+\dots+\lambda_N+0-(N-1)\lambda_1+\lambda_2+\dots + \lambda_N \\
&=2(\lambda_1+\lambda_2+\dots+\lambda_N)
\end{split}
\label{vl}\end{equation}
This is minimized by taking $(\lambda_1,\lambda_2,\dots,\lambda_{N})=(1,0,\dots,0)$, which gives $v\cdot \lambda'=2$. Thus, the fixture contains no free hypermultiplets. It is easy to check that the global symmetry is enhanced to $Sp(2N)_{2N+2} \times U(1)$ by the index 1 embedding

\begin{displaymath}
\begin{split}
Sp(2N)&\supset Sp(N) \times Sp(N) \\
\mathbf{Adj} &=(\mathbf{Adj,1})+(\mathbf{1,Adj})+(\mathbf{2N,2N})
\end{split}
\end{displaymath}
and the level of $Sp(2N)$ therefore saturates the unitarity bound in \cite{Beem:2013sza}.

With $n=0$, the S-duality is then given by \eqref{Sduality}, which implies that this SCFT has an effective number of vector and hypermultiplets given by $(n_h,n_v)=((2N+1)^2, (2N+1)^2-1-N(2N+1))$, which encode the trace anomaly coefficients $(a,c)=(\frac{n_h+5n_v}{24},\frac{n_h+2n_v}{12})=(\frac{1+19N+14N^2}{24},\frac{1+10N+8N^2}{12})$. We see that the $4D$ central charge is given by the Sugawara construction:

\begin{equation}
\begin{split}\label{sugawaraGen}
c_{4D}&=-\frac{1}{12}\left(\frac{-(N+1)2N(4N+1)}{-(N+1)+2N+1}+1\right) \\
&=\frac{1+10N+8N^2}{12}.
\end{split}
\end{equation}
It is easy to check that the saturation of these unitarity bounds follows from chiral ring relations at order $\tau^4$ in the $\mathbf{1}_0+\mathbf{(\frac{1}{2}(4N+1)(4N-2))_0}$ of $Sp(2N) \times U(1)$ by taking the plethystic log of \eqref{SCIhigher}. 

A similar analysis can be carried-through for the $R_{2,2N-1}$ theories. As already pointed out in \cite{Beem:2013sza}, the level of the $Spin(4N+2)_{4N-2}$ saturates the unitarity bound and the stress tensor of the 2D chiral algebra takes the Sugawara form, \eqref{sugawara}.

\section*{Acknowledgements}\label{Acknowledgements}
\addcontentsline{toc}{section}{Acknowledgements}
We would like to thank Yuji Tachikawa and Noppadol Mekareeya for helpful discussions. The work of J.D. and A.T. was supported in part by the National Science Foundation under Grant No. PHY-1316033. The work of O.C. was supported in part by the INCT-Matem\'atica and the ICTP-SAIFR in Brazil through a Capes postdoctoral fellowship. O.C. would like to thank the Johns Hopkins University, and especially Jared Kaplan, for hospitality while this work was being completed. A.T. thanks the hospitality of the Kavli IPMU at the University of Tokyo where part of this work was completed under NSF EAPSI award number IIA-1413868 and Japan Society for the Promotion of Science (JSPS) Summer Program 2014, and the Yukawa Institute for Theoretical Physics at Kyoto University for hospitality during the workshop YITP-W-14-4 ``Strings and Fields". He would especially like to thank Yuji Tachikawa and Yu Nakayama for their kind hospitality during his stay in Japan.

\bibliographystyle{utphys}
\bibliography{ref}

\end{document}